\documentclass{article}
\usepackage{graphicx}
\usepackage{amsmath, amssymb}
\usepackage{booktabs}
\usepackage{tabularx}
\usepackage{float}     
\usepackage{placeins}  
\usepackage{array}     
\usepackage{adjustbox}
\usepackage[round]{natbib}
\usepackage{graphicx}
\newcolumntype{Y}{>{\raggedright\arraybackslash}X}

\newcolumntype{Y}{>{\raggedright\arraybackslash}X}
\title{A Co-evolutionary Approach for Heston Calibration}
\author{
    Julian Gutierrez${}^{1}$ \\[1ex]
    {\small ${}^{1}$Department of Mathematics, Dartmouth College, Hanover, NH, USA}
}

\date{November 2025}

\begin{document}

\maketitle

\begin{abstract}
    We evaluate a co-evolutionary calibration framework for the Heston model in which a genetic algorithm (GA) over parameters is coupled to an evolving neural inverse map from option surfaces to parameters. While GA-history sampling can reduce training loss quickly and yields strong in-sample fits to the target surface, learning-curve diagnostics show a widening train--validation gap across generations, indicating substantial overfitting induced by the concentrated and less diverse dataset. In contrast, a broad, space-filling dataset generated via Latin hypercube sampling (LHS) achieves nearly comparable calibration accuracy while delivering markedly better out-of-sample stability across held-out surfaces. These results suggest that apparent improvements from co-evolutionary data generation largely reflect target-specific specialization rather than a more reliable global inverse mapping, and that maintaining dataset diversity is critical for robust amortized calibration.
\end{abstract}

\noindent \textbf{Keywords:} Heston model calibration, neuro-evolution, dataset, Latin Hypercube Sampling, Inverse neural surrogate

\section{Introduction}
Modern options trading and risk management require model calibration that is both accurate and fast. In production settings, parameters must be updated routinely---often daily and sometimes intraday---across many underlyings and maturities, under tight latency constraints. A calibration routine that is numerically fragile or too slow can negate the practical value of an otherwise expressive pricing model \citep{BuchelEtAl2022,LiuEtAl2019}.

The classical Black--Scholes framework \citep{BlackScholes1973} remains foundational due to its analytical tractability and physical interpretability, yet its diffusion-only dynamics with constant volatility do not reproduce the systematic skew and curvature observed in implied-volatility surfaces. This empirical mismatch has motivated richer parametric models that capture these features, including stochastic volatility and jumps in returns \cite{Heston1993,CarrEtAl2002,HorvathEtAl2021}.

Despite their improved fit, these models shift the central difficulty from pricing to calibration. Fitting a stochastic-volatility model entails solving a high-dimensional inverse problem that is typically nonconvex and computationally expensive, since each objective evaluation requires pricing an entire surface across strikes and maturities via numerical integration. As a result, calibration can become the dominant bottleneck in real-time pipelines \citep{GilliSchumann2011,CuiEtAl2017,LiuEtAl2019}.

Calibration is typically posed as minimizing the difference between market option quotes and model prices over a grid of strikes and maturities. For stochastic-volatility models---especially Heston-type specifications---this objective is generally nonconvex and exhibits multiple local minima, making optimizer choice consequential \citep{GilliSchumann2011,OrtizEtAl2022,LiuEtAl2019}. One class of approaches utilize global optimizers---simulated annealing, particle swarm optimization, differential evolution, and related evolutionary algorithms---which do not require derivatives and tend to be less sensitive to initialization \citep{GilliSchumann2010,GilliSchumann2011,MugandaKasamani2023,HaringHochreiter2015}. The drawback is computational: these methods often need a large number of iterations and objective evaluations, and the burden grows quickly with model dimension, which is problematic when calibrating multi-parameter dynamics to a full surface. A second strand uses gradient schemes, running a local optimizer from a set of initial points and selecting the best result; these can be substantially faster than global search when only a small number of starts is needed \citep{CuiEtAl2017,OrtizEtAl2022}. However, gradient-based methods often struggle because the objective is not smoothly differentiable or gradients are not readily available, so one must resort to finite-difference approximations that can be expensive and inaccurate. The Heston family is a representative case where both prices and sensitivities are often obtained through inverse-Fourier or other numerical integration procedures, so gradient computation can be noisy, and the integrands may become discontinuous or highly oscillatory for certain parameter combinations---precisely the regimes that can occur during search \citep{CuiEtAl2017,ZhangEtAl2024}.

Recent work has increasingly shifted toward machine-learning-based calibration to bypass several of these numerical difficulties. The common idea is to move expensive computation to an offline phase and deploy a learned mapping online: once a network is trained, inference is essentially instantaneous and can be executed at scale across many underlyings and surfaces. This amortized workflow is especially attractive in trading contexts because training can be performed before the market opens, while the calibrated parameters produced at runtime can be fed directly into pricing and risk systems with minimal latency \citep{LiuEtAl2019,HorvathEtAl2021,BuchelEtAl2022}. Moreover, neural surrogates can reduce reliance on fragile numerical derivatives by replacing---or serving as a warm-start for---gradient-based iterative solvers with a direct approximation of the inverse map from an observed surface to model parameters \citep{LiuEtAl2019,ZhangEtAl2024}.

A central practical question in these approaches is how to construct a training dataset that is both informative and efficient. A common strategy is to generate synthetic training examples by sampling the model's parameter space using space-filling designs such as Latin hypercube sampling (LHS), pricing a large collection of surfaces, and training a network to learn the inverse mapping \citep{LiuEtAl2019}. While this is conceptually straightforward, it introduces an implicit modeling choice: it is not obvious which parameter ranges should be sampled, how wide the sampling domain should be, or whether it is preferable to concentrate samples near the market-relevant region rather than covering the parameter space broadly \citep{HorvathEtAl2021}. In other words, the success of amortized calibration depends not only on network architecture, but also on a prior over parameters encoded by the training distribution.

To investigate this trade-off, we introduce a dual-population co-evolutionary calibration framework that couples parameter search with neuro-evolution of an inverse mapping. At each generation, a genetic algorithm maintains a population of candidate parameter vectors and identifies an elite subset based on calibration error to the target market surface. The elite parameters are then used to generate model-implied option surfaces, and these elite (surface, parameter) pairs define the incremental training set for a simultaneously evolving population of neural networks. Each network implements an inverse map from a flattened option surface to the model parameter vector, and evolves both in weights and architecture through crossover and mutation operators. Network fitness is evaluated using a direct calibration score obtained by pricing the target surface using the network's predicted parameters. The best-performing networks are then used to inject new candidate parameters into the GA population by evaluating the networks on the target surface and adding small perturbations to their predictions, replacing the weakest GA individuals. This establishes a closed feedback loop in which GA elites define the training signal for the inverse networks, and the inverse networks, in turn, provide targeted proposals that accelerate subsequent global search.

Empirically, this framework allows us to test the often unstated assumption that ``sampling closer to the true parameters'' yields a better inverse model than broad coverage. In our experiments, we find that inverse networks trained on LHS-generated datasets can outperform networks trained solely on GA-history datasets, indicating that diversity in parameter-space coverage is critical for learning a stable inverse map. We also find that NN-based seeding improves the convergence speed of the GA relative to a plain GA baseline, but this acceleration alone does not compensate for the loss in training diversity when the dataset is generated purely from optimizer history. These results suggest that while market-guided sampling is appealing, inverse calibration benefits from training distributions that remain sufficiently space-filling, and they motivate hybrid strategies that retain broad coverage while still leveraging optimizer feedback.

\section{The Heston stochastic-volatility model}

\subsection{Risk-neutral dynamics and parameters}
In the Heston model \citep{Heston1993} we work under a risk-neutral measure \(\mathbb{Q}\) with constant risk-free rate \(r\). In the Heston model, the underlying price \(S_t\) and its instantaneous variance \(v_t\) jointly evolve as a coupled diffusion with correlated Brownian shocks:
\begin{equation}
\begin{aligned}
dS_t &= r S_t\,dt + \sqrt{v_t}\,S_t\, dW_{1,t},\\
dv_t &= \kappa(\lambda - v_t)\,dt + \sigma \sqrt{v_t}\, dW_{2,t},\\
\mathbb{E}^{\mathbb{Q}}[dW_{1,t}\,dW_{2,t}] &= \rho\,dt.
\end{aligned}
\end{equation}
The parameter vector governing the variance process is
\begin{equation}
\theta_H := (\kappa,\lambda,\sigma,\rho,v_0),
\end{equation}
where \(\kappa>0\) controls mean reversion, \(\lambda>0\) is the long-run variance level, \(\sigma>0\) is the vol-of-vol, \(\rho\in[-1,1]\) is the return--variance correlation, and \(v_0>0\) is the initial variance.

A practical consideration (both for numerical stability and data generation) is to enforce constraints that keep the variance process well-behaved; a common sufficient condition for strict positivity is the Feller constraint \(2\kappa\lambda>\sigma^2\) \citep{CoxIngersollRoss1985, Feller1951}.

\subsection{Semi-analytical pricing via Fourier inversion}
For calibration, we require fast evaluation of model prices across a grid of strikes and maturities. Heston admits a semi-analytical pricing expression for European calls based on Fourier inversion. Let \(\tau\) denote time to maturity and \(K\) the strike, and define the full input tuple
\begin{equation}
\theta := (\kappa,\lambda,\sigma,\rho,v_0,S_0,r,\tau,K).
\end{equation}
Following \cite{ZhangEtAl2024} then the time-0 call price can be written as
\begin{equation}
C(\theta)= S_0\Pi_1 - K e^{-r\tau}\Pi_2,
\end{equation}
with
\begin{equation}
\Pi_1 = \frac{1}{2}+\frac{1}{\pi}\int_{0}^{\infty}\Re\!\left(\frac{\varphi_\tau(u-i)}{iu}e^{-iuk}\right)\,du,
\qquad
\Pi_2 = \frac{1}{2}+\frac{1}{\pi}\int_{0}^{\infty}\Re\!\left(\frac{\varphi_\tau(u)}{iu}e^{-iuk}\right)\,du,
\end{equation}
where \(k=\ln K\) and \(\varphi_\tau(u)\) is the characteristic function of \(\ln(S_\tau)\). Put prices follow immediately from put--call parity.

Because calibrations repeatedly evaluate these integrals, the specific characteristic-function representation matters: formulas can suffer from branch-cut instabilities due to complex logarithms. A numerically stable specification is
\begin{equation}
\varphi_\tau(u)=\exp\!\big(C_\tau(u) + D_\tau(u)\,v_0 + iu\ln S_0\big),
\end{equation}
with
\begin{equation}
\begin{aligned}
C_\tau(u) &= iru\tau + \frac{\kappa\lambda}{\sigma^2}\left((\kappa-i\rho\sigma u-d(u))\tau -2\ln\!\left(\frac{1-g(u)e^{-d(u)\tau}}{1-g(u)}\right)\right),\\
D_\tau(u) &= \frac{\kappa-i\rho\sigma u-d(u)}{\sigma^2}\left(\frac{1-e^{-d(u)\tau}}{1-g(u)e^{-d(u)\tau}}\right),\\
g(u) &= \frac{\kappa-i\rho\sigma u-d(u)}{\kappa-i\rho\sigma u+d(u)},\qquad
d(u)=\sqrt{(\kappa-i\rho\sigma u)^2+\sigma^2(iu+u^2)}.
\end{aligned}
\end{equation}

\subsection{Calibration objective}
Given market quotes over a strike--maturity grid \(\{(K_m,\tau_m)\}_{m=1}^M\), calibration is posed as an inverse problem: find \(\theta_H\) minimizing a pricing-misfit objective, typically least squares over observed prices,
\begin{equation}
\theta_H^\star \in \arg\min_{\theta_H\in\Theta_H}\frac{1}{M}\sum_{m=1}^M\Big(C(\theta_H;S_0,r,\tau_m,K_m)-C^{\text{mkt}}_{m}\Big)^2,
\end{equation}
subject to our feasibility constraints. 

\section{Co-evolutionary approach}

Our calibration method couples a global parameter search with an evolving inverse map from option surfaces to Heston parameters. The algorithm maintains two populations that interact through a feedback loop: (i) a genetic algorithm (GA) over Heston parameter vectors and (ii) a population of neural networks trained and evolved to predict parameters from a target surface.

\subsection{Population structure}
At generation \(g\), the GA population is
\begin{equation}
\mathcal{P}_{\mathrm{GA}}^{(g)}=\{\theta_1^{(g)},\theta_2^{(g)},\dots,\theta_N^{(g)}\},\qquad
\theta_n^{(g)}\in\mathbb{R}^5,
\end{equation}
where \(\theta=(\kappa,\lambda,\sigma,\rho,v_0)\) denotes the Heston parameter vector. In parallel, we maintain a neural population
\begin{equation}
\mathcal{P}_{\mathrm{NN}}^{(g)}=\{\mathrm{NN}_1^{(g)},\mathrm{NN}_2^{(g)},\dots,\mathrm{NN}_M^{(g)}\}.
\end{equation}
Each network implements an inverse mapping
\begin{equation}
f_i:\mathbb{R}^{KT}\to\Theta,\qquad \Theta\subset\mathbb{R}^5,
\end{equation}
taking as input a flattened option-price surface on a \(K\times T\) strike--maturity grid and outputting a parameter proposal in \(\Theta\), the feasible Heston parameter space.

The GA population \(\mathcal{P}_{\mathrm{GA}}^{(g)}\) is advanced by elitist selection followed by variation in the feasible parameter space \(\Theta\). In each generation, the top fraction \(\varepsilon_{\mathrm{GA}}\) of parameter vectors are retained and forms the parent pool for reproduction. Offspring are generated by selecting two parents from this elite pool, applying crossover, and then applying mutation.

\begin{table}[H]
\caption{Genetic-algorithm operators for Heston parameter evolution.}
\label{tab:ga-operators}
\small
\begin{adjustbox}{center, width=\textwidth }
\begin{tabular}{@{}llll@{}}
\toprule
Operator & Probability & Choice set / rule & Constraint \\ \midrule
Selection (elitism)
& ---
& Preserve the top \(\varepsilon_{\mathrm{GA}}\) fraction by fitness
& \(\varepsilon_{\mathrm{GA}}\in(0,1)\). \\

Crossover (arithmetic) 
& \(p_{\mathrm{x,GA}}\)
& Select \(\theta_{p1},\theta_{p2}\) from the elite pool and set
\(\theta_{\text{child}}=\tfrac{1}{2}(\theta_{p1}+\theta_{p2})\).
& Parents drawn from elites. \\

Mutation (Gaussian)
& \(p_{\mathrm{m,GA}}\)
& Perturb \(\theta\) as \(\theta'=\theta+\xi\), \(\xi\sim\mathcal{N}(0,\sigma_{\mathrm{mut}}^2 I)\).
& Enforce \(\theta'\in\Theta\) after mutation. \\
\bottomrule
\end{tabular}
\end{adjustbox}
\end{table}

\subsection{Network representation and variation}
Each network \(\mathrm{NN}_i\) is defined by an architecture and its weights. We encode the architecture as
\begin{equation}
A_i=(L_i,H_i,\alpha_i),
\end{equation}
where \(L_i\) is the number of hidden layers, \(H_i=(h_{i,1},\dots,h_{i,L_i})\) are hidden widths, and \(\alpha_i\) is the activation type. The trainable parameters are
\begin{equation}
W_i=\{(W_{\ell},b_{\ell})\}_{\ell=1}^{L_i+1},
\end{equation}
where \(W_{\ell}\) and \(b_{\ell}\) denote the weight matrix and bias vector of layer \(\ell\). Let \(d_0=KT\) be the input dimension, let \(d_{\ell}=h_{i,\ell}\) for \(\ell=1,\dots,L_i\) be the hidden widths, and let \(d_{L_i+1}=5\) be the output dimension. Then
\begin{equation}
W_{\ell}\in\mathbb{R}^{d_{\ell}\times d_{\ell-1}},\qquad
b_{\ell}\in\mathbb{R}^{d_{\ell}},\qquad \ell=1,\dots,L_i+1.
\end{equation}

Networks evolve through (i) continuous variation of weights and (ii) occasional architecture changes. For weight evolution, we use arithmetic crossover and Gaussian mutation,
\begin{equation}
W_{\text{child}}=\tfrac{1}{2}(W_{p1}+W_{p2}),\qquad
W' = W + \Delta W,\ \ \Delta W\sim\mathcal{N}(0,\sigma_{\mathrm{w}}^2 I).
\end{equation}

\begin{table}[H]
\caption{Neural-network architecture search space and mutation operators.}
\label{tab:arch-mutations}
\small
\begin{adjustbox}{center, width=\textwidth}
\begin{tabular}{@{}llll@{}}
\toprule
Operator & Probability & Choice set / rule & Constraint \\ \midrule
Add hidden layer & \(p_{\mathrm{add}}\) & \(h_{\text{new}}\in\{32,64,128,256\}\) & inserted if feasible \\
Remove hidden layer & \(p_{\mathrm{rm}}\) & remove a randomly chosen layer & requires \(L_i>1\) \\
Modify layer width & \(p_{\mathrm{mod}}\) & \(h_{i,\ell}\leftarrow h_{\text{new}},\ h_{\text{new}}\in\{16,32,64,128,256,512\}\) & any \(\ell\in\{1,\dots,L_i\}\) \\
Change activation & \(p_{\alpha}\) & \(\alpha_i \leftarrow \alpha_{\text{new}},\ \alpha_{\text{new}}\in\{\mathrm{ReLU},\mathrm{Tanh},\mathrm{LeakyReLU},\mathrm{ELU}\}\) & --- \\
Architecture crossover & --- & \(L_{\text{child}}=\max(L_{p1},L_{p2})\) & hybridize widths/activation \\
\bottomrule
\end{tabular}
\end{adjustbox}
\end{table}

When two parents have incompatible depth/width, we construct a hybrid offspring by taking
\begin{equation}
L_{\text{child}}=\max(L_{p1},L_{p2}),\qquad
h_{\ell,\text{child}}=\left\lfloor\frac{h_{\ell,p1}+h_{\ell,p2}}{2}\right\rfloor,
\end{equation}
and selecting the child's activation function, \(\alpha_{\text{child}}\), uniformly from the parents' activations.

\subsection{Feedback loop: data generation, training, and injection}
Let \(\mathcal{E}_g\subset\{1,\dots,N\}\) denote the elite GA indices at generation \(g\) by calibration error against the target market surface). For each elite \(\theta_j^{(g)}\), we compute the corresponding model-implied option surface \(S(\theta_j^{(g)})\), flatten it to \(s_j^{(g)}=\mathrm{flat}(S(\theta_j^{(g)}))\in\mathbb{R}^{KT}\), and form the incremental dataset
\begin{equation}
\mathcal{D}_g=\{(s_j^{(g)},\theta_j^{(g)}): j\in\mathcal{E}_g\}.
\end{equation}
Each network is updated on \(\mathcal{D}_g\) using a squared-error objective,
\begin{equation}
\min_{W_i}\ \sum_{(s,\theta)\in\mathcal{D}_g}\left\|\mathrm{NN}_i(s;W_i)-\theta\right\|_2^2,
\end{equation}
implemented via a fixed number of gradient steps per generation.

Network fitness is assessed using (i) this surrogate prediction loss on \(\mathcal{D}_g\) and (ii) a direct calibration score on the target surface. Specifically, given the observed target surface \(S_{\text{target}}\),
\begin{equation}
\hat{\theta}_i = \mathrm{NN}_i(\mathrm{flat}(S_{\text{target}})),
\end{equation}
and the direct score is the calibration loss obtained by pricing with \(\hat{\theta}_i\) and comparing to the target grid under the same metric used for GA evaluation.

Finally, the best-performing networks propose new GA candidates by evaluating the target surface and adding exploration noise:
\begin{equation}
\theta_{\text{inject}}=\hat{\theta}_i+\zeta,\qquad \zeta\sim\mathcal{N}(0,\sigma_{\text{inj}}^2 I).
\end{equation}
These injected candidates replace the weakest GA individuals, while the remainder of the GA population is advanced by standard elitist selection, arithmetic crossover, and Gaussian mutation,
\begin{equation}
\theta_{\text{child}}=\tfrac{1}{2}(\theta_{p1}+\theta_{p2}),\qquad
\theta'=\theta+\xi,\ \ \xi\sim\mathcal{N}(0,\sigma_{\mathrm{mut}}^2 I).
\end{equation}
GA elites generate training signal for the inverse networks, and the networks return targeted proposals that accelerate subsequent global search.

\begin{figure}[H]
\centering
\includegraphics[width=\textwidth]{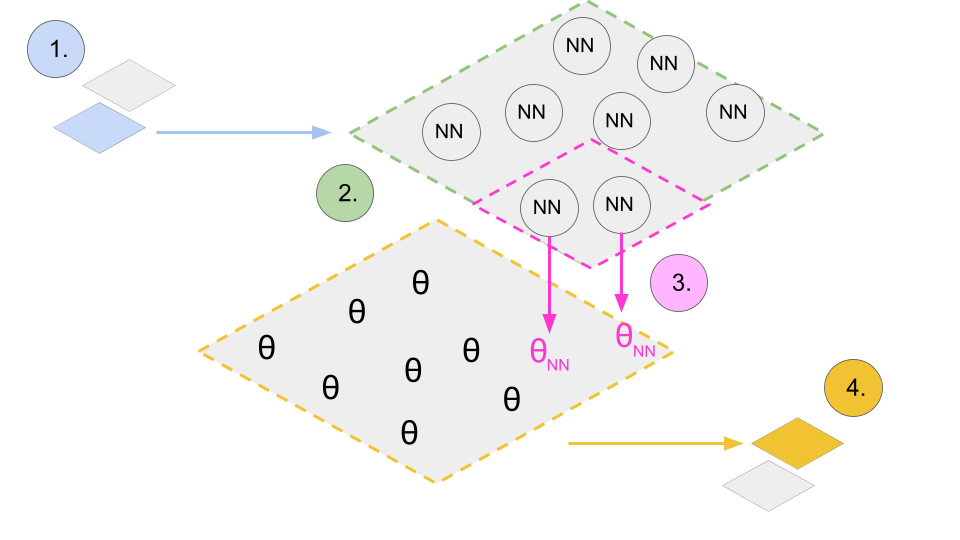}

\includegraphics[width=\textwidth]{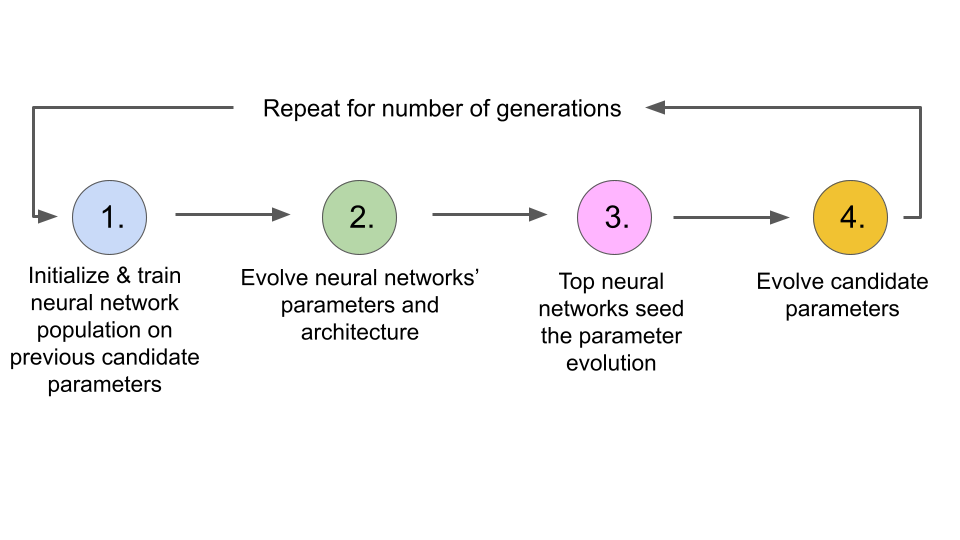}

\caption{Overview of the co-evolutionary calibration loop. A population of neural networks is trained and evolved using surfaces generated by elite GA candidates; top networks then propose parameter seeds that are injected into the GA, accelerating subsequent parameter evolution. The process repeats for a fixed number of generations.}
\label{fig:coev-overview}
\end{figure}

\section{Empirical results}
We first evaluate the proposed co-evolutionary calibration method in a fully controlled synthetic setting, where the ground-truth Heston parameters are known and performance can be measured directly. The goals of these experiments are (i) to compare convergence behavior against a plain genetic algorithm baseline under a fixed compute budget, (ii) to quantify time-to-accuracy relative to a common local optimizer (LBFGS), and (iii) to test how the induced training-data distribution affects the performance and generalization of the learned inverse map. In particular, we compare inverse networks trained on (a) the concentrated dataset produced by the co-evolutionary loop and (b) a broad space-filling dataset generated via Latin hypercube sampling (LHS), isolating the role of dataset diversity versus target-focused sampling.

\subsection{Experimental design and synthetic data generation}\label{sec:synthetic}

Synthetic option price surfaces are generated by first sampling a parameter vector
\(\theta^\star=(\kappa,\lambda,\sigma,\rho,v_0)\) at random from the feasible Heston parameter space \(\Theta\subset\mathbb{R}^5\). For each sampled \(\theta^\star\), we price European options on a fixed strike--maturity grid of size \(K\times T\) using the semi-analytical Heston pricing engine described in Section~2, producing a surface
\(S(\theta^\star)\in\mathbb{R}^{K\times T}\). The corresponding flattened representation
\(s^\star=\mathrm{flat}(S(\theta^\star))\in\mathbb{R}^{KT}\) serves as the observation provided to inverse networks and as the target surface for calibration.
\begin{table}[H]
\centering
\caption{Ranges of the Heston model parameters in Eq. (1), values come from \cite{ZhangEtAl2024}}
\label{tab:synthetic-ranges}
\begin{tabular}{@{}cc@{}}
\toprule
Parameter & Range \\ \midrule
\(\kappa\) & \([\,0.005,\ 5\,]\) \\
\(\lambda\) & \([\,0,\ 1\,]\) \\
\(\sigma\) & \([\,0.1,\ 1\,]\) \\
\(\rho\) & \([\, -0.95,\ 0\,]\) \\
\(v_0\) & \([\,0,\ 1\,]\) \\
\(r\) & \([\,0,\ 0.10\,]\) \\
\(\tau\) & \([\,0.05,\ 1\,]\) \\
\bottomrule
\end{tabular}
\end{table}

\subsection{Hyperparameters}\label{sec:hyperparams}
The synthetic experiments use fixed hyperparameters across all trials to isolate the effect of the proposed seeding mechanism. We group settings by (i) GA evolution of \(\theta\)-candidates, (ii) neuro-evolution, and (iii) neural-network training.

\begin{table}[H]
\centering
\caption{Hyperparameters used in the synthetic experiments by component.}
\label{tab:hyperparams}
\small
\begin{adjustbox}{center, width=\textwidth -2 cm}
\begin{tabular}{@{}llcl@{}}
\toprule
Component & Hyperparameter & Value & Notes \\ \midrule
\multicolumn{4}{@{}l}{\textit{Genetic algorithm (GA)}}\\
& \(N\) & 50 & Number of \(\theta\)-candidates per generation. \\
& \(G\) & 10 & Same budget for GA and co-evolutionary runs. \\
& \(\varepsilon_{\mathrm{GA}}\) & 0.2 & Defines elite parent pool used for crossover. \\
& \(\mu_{\mathrm{GA}}\) & 0.1 & Per-parameter mutation probability. \\
& \(p_{\mathrm{x,GA}}\) & 0.3 \\
& \(p_{\mathrm{m,GA}}\) & 0.2 \\
& \(\sigma_{\mathrm{mut}}\) & \(0.05\times\) range & Gaussian noise scale as a fraction of each parameter's box range. \\ \midrule

\multicolumn{4}{@{}l}{\textit{Neuro-evolution (NN population + GA seeding)}}\\
& \(M\) & 20 & Number of networks evolved in parallel. \\
& \(\varepsilon_{\mathrm{NN}}\) & 0.2 & Survival fraction in the NN population. \\
& \(\varepsilon_{\mathrm{inj}}\) & 0.2 & Fraction of GA population replaced by NN-proposed seeds. \\
& \(\mu_{\mathrm{w}}\) & 0.1 & Probability of perturbing each weight. \\
& \(\sigma_{\mathrm{w}}\) & 0.02 & Std.\ dev.\ of Gaussian weight perturbations. \\
& \(\sigma_{\mathrm{inj}}\) & 0.01 & Noise added to NN parameter predictions before injection. \\
& \(p_{\mathrm{arch}}\) & 0.05  \\
& \(p_{\mathrm{rm}}\) & 0.3 \\ 
& \(p_{\mathrm{add}}\) & 0.3 \\ 
& \(p_{\mathrm{mod}}\) & 0.5 \\ 
& \(p_{\alpha}\) & 0.2 &  \\
\midrule

\multicolumn{4}{@{}l}{\textit{Neural-network training (per-generation updates)}}\\
& Optimization algorithm & Adam \\
& Epochs per generation & 5 & Adam steps on the current training set. \\
& Initial learning rate & 0.001\\
& Decay rate & 0.9 \\
& Loss & MSE & Parameter-space regression loss. \\
& Feedback epochs & 2 & Retraining on the latest GA-elite data. \\
& Feedback frequency & 1 & Retrain every generation. \\
& Training/Test set ratio & 7:3 \\ 
& Batch size & 64 \\
\bottomrule
\end{tabular}
\end{adjustbox}
\end{table}

\subsection{Results}
\subsubsection{Convergence comparison}
\noindent We first compare convergence behavior on synthetic surfaces. We plot the calibration loss (RMSE) as a function of generation for (i) a plain GA evolving \(\theta\)-candidates directly and (ii) the proposed co-evolutionary method, in which the NN population proposes parameter seeds that are injected into the GA. The co-evolutionary curve descends more rapidly in the early generations and continues improving after the plain GA begins to plateau. This indicates that NN-driven injections provide informative proposals that accelerate global search and help sustain progress over a longer horizon under the same evolutionary budget.

\begin{figure}[H]
\centering
\includegraphics[width=\textwidth]{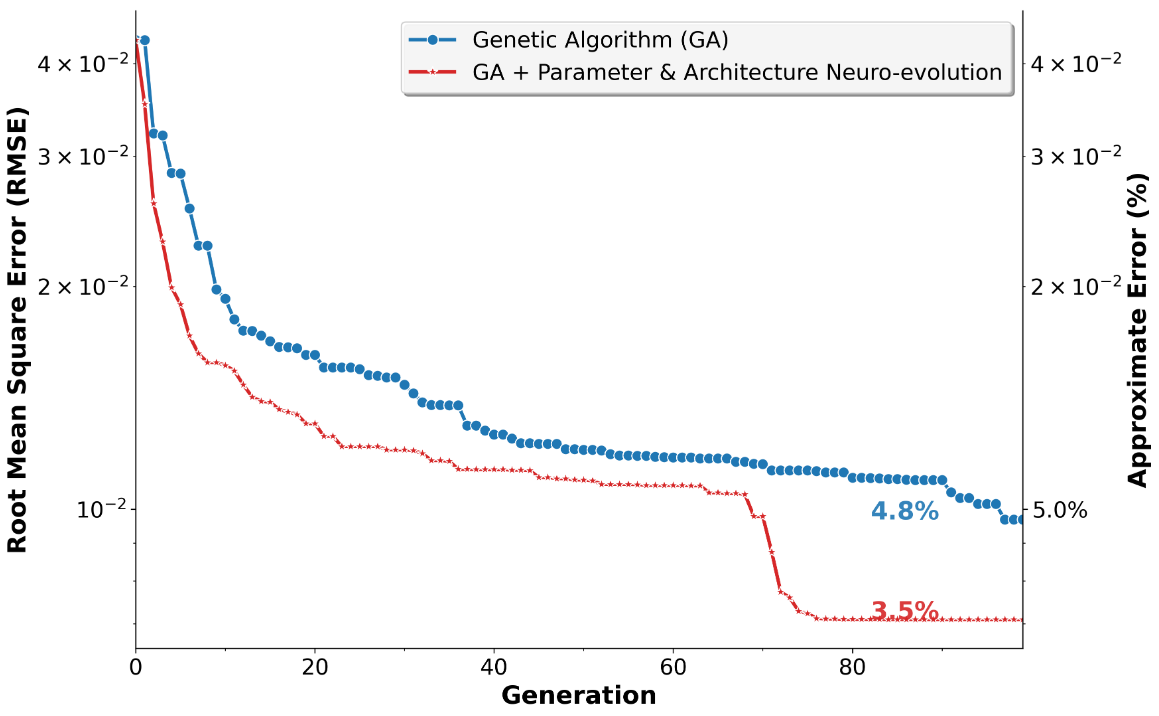}
\caption{Convergence on synthetic calibration targets. Calibration RMSE versus generation for a plain genetic algorithm (GA) baseline and the proposed co-evolutionary method (GA + parameter \& architecture neuro-evolution). Under the same compute budget, the co-evolutionary method reduces error more quickly in early generations and continues improving after the GA plateaus, consistent with NN-based seeding accelerating and stabilizing the global search.}
\label{fig:conv-ga-vs-coev}
\end{figure}
\subsubsection{Time to convergence distribution}

\noindent To benchmark against a representative gradient-based local method, we additionally compare our evolutionary approach to LBFGS. For each synthetic target, we run LBFGS under the same pricing objective to obtain a reference calibration error level. We then define the \emph{time-to-threshold} (TTT) as the smallest GA generation \(g\) at which the evolutionary method first achieves an error no worse than this LBFGS reference. Repeating this procedure across \(n\) independent synthetic targets yields a distribution of TTT values, which summarizes how many evolutionary generations are typically required to match the accuracy of a standard gradient routine.

\begin{figure}[H]
\centering
\includegraphics[width=\textwidth]{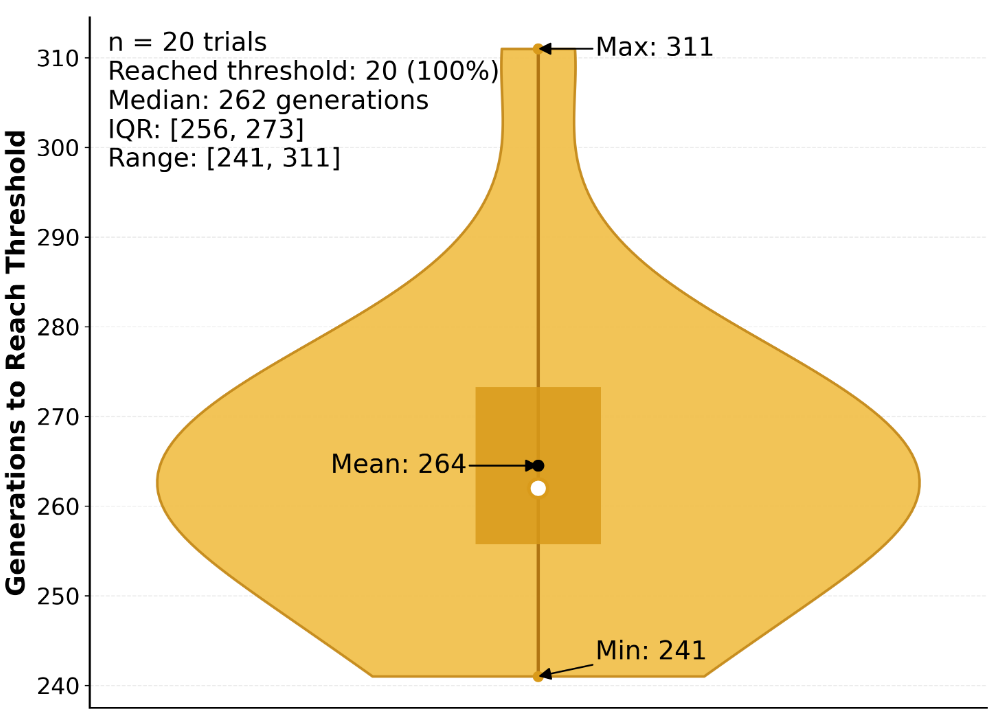}
\caption{Time-to-threshold relative to LBFGS on synthetic targets. The reported time-to-threshold is the first GA generation at which the evolutionary method matches that reference level RMSE. The distribution over \(n=20\) trials indicates that matching LBFGS-level accuracy typically requires around 264 generations to match the perforamnce of LBFGS.}
\label{fig:ttt-lbfgs}
\end{figure}

\begin{table}[H]
\centering
\caption{Summary of neural-architecture statistics over training generations. Reported values are computed over the NN population at each checkpoint.}
\label{tab:nn-arch-summary}
\small
\begin{adjustbox}{center, width=\textwidth -2 cm}
\begin{tabular}{@{}r r r r r r l l l r@{}}
\toprule
Generation & Avg layers & Avg nodes & Std nodes & Min nodes & Max nodes & Most common arch. & Frequency & Primary act. & Act. div. \\ \midrule
20  & 1.95 & 192.0 & 56.3 & 32 & 384 & \([128,64]\) & 17/20 & ReLU & 2 \\
40  & 1.95 & 179.2 & 37.0 & 64 & 256 & \([128,64]\) & 14/20 & ReLU & 1 \\
60  & 2.10 & 184.0 & 52.5 & 64 & 320 & \([128,64]\) & 11/20 & ReLU & 2 \\
80  & 2.05 & 188.8 & 75.0 & 64 & 320 & \([128,64]\) & 9/20  & ReLU & 4 \\
100 & 2.15 & 208.0 & 78.7 & 64 & 384 & \([128,64]\) & 6/20  & ReLU & 3 \\
\bottomrule
\end{tabular}
\end{adjustbox}
\end{table}

The architecture-summary table suggests that the evolutionary pressure is not only shrinking loss but also gradually favoring higher-capacity models. The average depth increases from 1.95 layers at generation 20 to 2.15 layers by generation 100, alongside an increase in average total nodes from 192 to 208 and a rising spread in node counts (Std Nodes from 56.3 to 78.7, with Max Nodes returning to 384). At the same time, the initially dominant architecture 
[128,64]
[128,64] becomes less ubiquitous (frequency declines from 17/20 to 6/20) and activation diversity increases from 1--2 up to 3--4, indicating a broader exploration of architectural variants later in evolution. Taken together, these trends are consistent with a slow drift toward deeper, more expressive networks as the incremental dataset grows, while the population becomes less concentrated around a single “default” design.
\begin{table}[H]
\centering
\caption{Representative evolved architectures at generation 20.}
\label{tab:nn-arch-samples-20}
\small
\begin{tabular}{@{}l l r r l@{}}
\toprule
NN ID & Architecture & Num layers & Total nodes & Activation \\ \midrule
NN-1 & \([128,64]\)    & 2 & 192 & ReLU \\
NN-2 & \([256,128]\)   & 2 & 384 & ReLU \\
NN-3 & \([32]\)        & 1 & 32  & ELU \\
NN-4 & \([32,128]\)    & 2 & 160 & ELU \\
NN-5 & \([64,32,32]\)  & 3 & 128 & ReLU \\
\bottomrule
\end{tabular}
\end{table}

\begin{table}[H]
\centering
\caption{Representative evolved architectures at generation 40.}
\label{tab:nn-arch-samples-40}
\small
\begin{tabular}{@{}l l r r l@{}}
\toprule
NN ID & Architecture & Num layers & Total nodes & Activation \\ \midrule
NN-1 & \([128,64]\)   & 2 & 192 & ReLU \\
NN-2 & \([64,64]\)    & 2 & 128 & ReLU \\
NN-3 & \([64]\)       & 1 & 64  & ReLU \\
NN-4 & \([128,32]\)   & 2 & 160 & ReLU \\
NN-5 & \([32,128]\)   & 2 & 160 & ReLU \\
\bottomrule
\end{tabular}
\end{table}

\begin{table}[H]
\centering
\caption{Representative evolved architectures at generation 60.}
\label{tab:nn-arch-samples-60}
\small
\begin{tabular}{@{}l l r r l@{}}
\toprule
NN ID & Architecture & Num layers & Total nodes & Activation \\ \midrule
NN-1 & \([128,64]\)     & 2 & 192 & ReLU \\
NN-2 & \([256]\)        & 1 & 256 & ReLU \\
NN-3 & \([64,32,128]\)  & 3 & 224 & ReLU \\
NN-4 & \([64]\)         & 1 & 64  & Tanh \\
NN-5 & \([32,128]\)     & 2 & 160 & ReLU \\
\bottomrule
\end{tabular}
\end{table}

\begin{table}[H]
\centering
\caption{Representative evolved architectures at generation 80.}
\label{tab:nn-arch-samples-80}
\small
\begin{tabular}{@{}l l r r l@{}}
\toprule
NN ID & Architecture & Num layers & Total nodes & Activation \\ \midrule
NN-1 & \([128,64]\)     & 2 & 192 & ReLU \\
NN-2 & \([64,32]\)      & 2 & 96  & ReLU \\
NN-3 & \([64,256]\)     & 2 & 320 & ReLU \\
NN-4 & \([64,64]\)      & 2 & 128 & ReLU \\
NN-5 & \([128,64,64]\)  & 3 & 256 & ReLU \\
\bottomrule
\end{tabular}
\end{table}

\begin{table}[H]
\centering
\caption{Representative evolved architectures at generation 100.}
\label{tab:nn-arch-samples-100}
\small
\begin{tabular}{@{}l l r r l@{}}
\toprule
NN ID & Architecture & Num layers & Total nodes & Activation \\ \midrule
NN-1 & \([128,64]\)      & 2 & 192 & ReLU \\
NN-2 & \([64,64]\)       & 2 & 128 & ReLU \\
NN-3 & \([128,32,32]\)   & 3 & 192 & ReLU \\
NN-4 & \([128,128,32]\)  & 3 & 288 & ReLU \\
NN-5 & \([128]\)         & 1 & 128 & ReLU \\
\bottomrule
\end{tabular}
\end{table}

\subsubsection{Neural network improvement over generations}

As the co-evolutionary loop progresses, the neural population is repeatedly retrained on an expanding set of GA-elite surface--parameter pairs, while its architecture is simultaneously perturbed by the mutation operators described in Section~3. This produces a moving target for learning: the dataset grows and shifts toward market-relevant regions, and the hypothesis class changes as depth, width, and activations evolve. To assess training quality across this process, we track the learning curves of representative networks at several generation checkpoints. The figure below reports the training and validation MSE as a function of epoch for checkpoints at generations 20, 40, 60, 80, and 100, illustrating how optimization dynamics and generalization behavior change as both the data stream and the architecture distribution evolve.

\begin{figure}[H]
\centering
\hspace{-1cm}
\includegraphics[width=\textwidth]{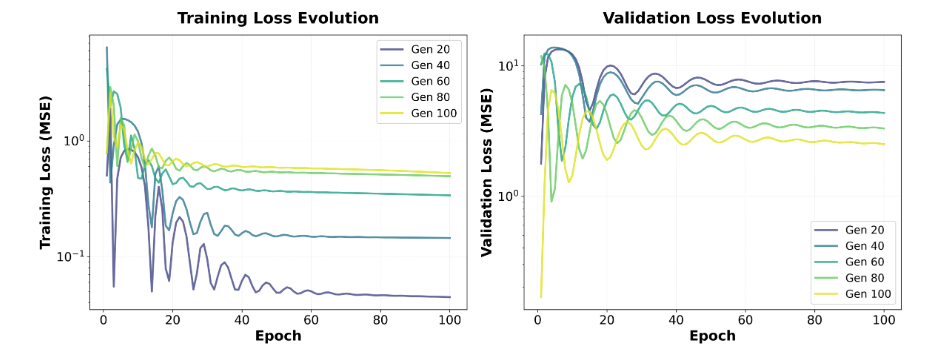}
\caption{Neural-network training quality across co-evolution. Training (left) and validation (right) MSE learning curves for checkpoint generations \(g\in\{20,40,60,80,100\}\). Curves reflect the combined effect of (i) an expanding GA-elite dataset and (ii) evolving NN architectures over generations.}
\label{fig:nn-loss-across-gens}
\end{figure}

\subsubsection{Overfitting analysis}

As the neural population accumulates additional GA-elite training pairs across generations, the training curves exhibit a clear downward shift in loss, while the corresponding validation curves improve much more slowly and remain materially higher. This widening train--validation gap is consistent with the inverse networks becoming increasingly specialized to the narrow region of parameter space emphasized by optimizer history, rather than learning an inverse map that generalizes across held-out surfaces. In other words, the same feedback loop that produces highly relevant training examples for seeding can also concentrate the data distribution, making it easier to reduce training error while providing limited coverage for out-of-sample validation---a pattern that motivates our subsequent overfitting analysis and the comparison against space-filling datasets.

\begin{figure}[H]
\centering
\includegraphics[width=\textwidth]{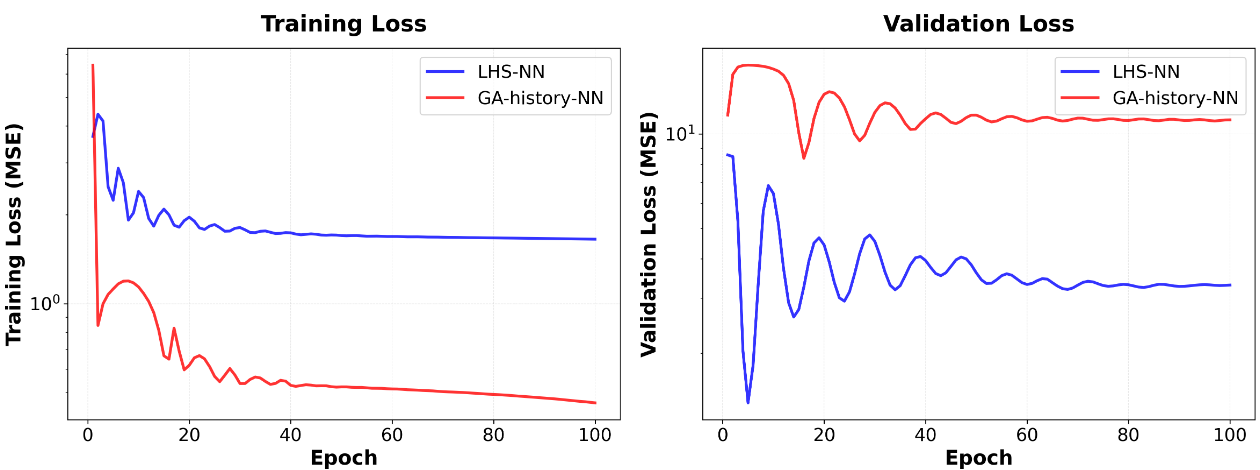}

\vspace{0.6em}

\includegraphics[width=\textwidth]{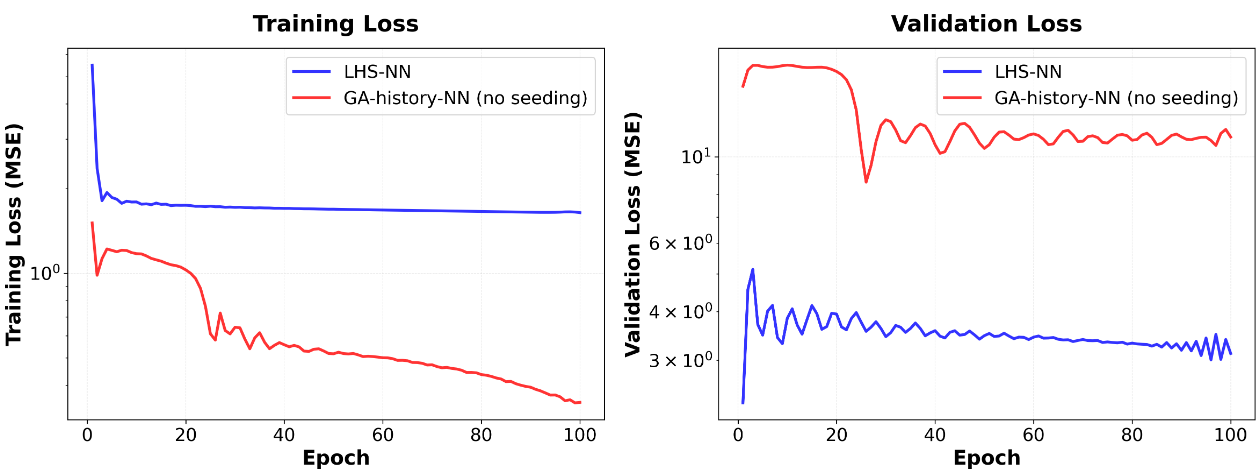}

\caption{Training/validation loss for inverse networks trained on a space-filling LHS dataset versus GA-history data. Top: seeding/warm-start enabled. Bottom: seeding disabled. In both settings, GA-history sampling yields low training loss but substantially higher validation loss, consistent with overfitting to the target surface induced by a concentrated and less diverse training distribution. LHS sampling, while not concentrated near the target, learns a more stable inverse mapping with improved generalization across held-out surfaces.}
\label{fig:lhs-vs-gahistory-seed-vs-noseed}
\end{figure}
\subsubsection{Real-surface calibration and LHS dataset comparison}\label{sec:lhs_real}

We next calibrate to an SPX option price surface, using the S\&P 500 index as the underlying (data from Yahoo Finance). This real-surface calibration provides the target surface used by the co-evolutionary loop, after which we compare two training-set constructions for the inverse network: (i) GA-history data collected from optimizer elites and (ii) a broad, space-filling dataset generated via Latin hypercube sampling, which isolates the effect of dataset diversity on generalization. In the LHS setting, we sample parameter vectors uniformly in the prescribed bounds and generate their corresponding model-implied surfaces under the same pricing engine and grid used throughout.

\begin{table}[H]
\centering
\caption{SPX real-surface calibration progress and parameter error over generations. Parameter columns report relative error (\%) for the co-evolutionary method.}
\label{tab:spx_calib_progress}
\small
\begin{adjustbox}{center, width=\textwidth}
\begin{tabular}{@{}rcccccc@{}}
\toprule
Generation & Loss & \(\kappa\) (\%) & \(\lambda\) (\%) & \(\sigma\) (\%) & \(\rho\) (\%) & \(v_0\) (\%) \\
\midrule
20  & 0.000298 & 400.607415 & 42.589554 & 17.772964 & 27.881316 & 25.709461 \\
40  & 0.000207 & 285.404704 & 38.461205 & 17.889609 & 27.070596 & 26.622294 \\
60  & 0.000139 & 153.712397 & 34.734711 & 21.746060 & 25.347386 & 16.781899 \\
80  & 0.000113 & 115.876770 & 31.543873 & 22.323575 & 24.976307 & \ \,6.892037 \\
100 & 0.000083 & \ \,58.191643 & 27.517073 & 22.514355 & 24.685334 & \ \,6.235404 \\
\bottomrule
\end{tabular}
\end{adjustbox}
\end{table}

\begin{table}[H]
\centering
\caption{U.S. Treasury rates used for the risk-free curve (weekly maturities).}
\label{tab:ust_rates}
\small
\begin{tabular}{@{}cc@{}}
\toprule
Maturity (weeks) & Rate (\%) \\
\midrule
4  & 4.24 \\
6  & 4.23 \\
13 & 4.23 \\
17 & 4.19 \\
26 & 4.07 \\
52 & 3.77 \\
\bottomrule
\end{tabular}
\end{table}

\begin{table}[H]
\centering
\caption{Summary statistics for the SPX option sample used in the real-surface experiments.}
\label{tab:spx_sample_summary}
\small
\begin{tabular}{@{}lccc@{}}
\toprule
Ticker & \# samples & Maturity (days) & Moneyness \(\ln(K/S_0)\) \\
\midrule
SPX & 152 & [3,\ 255] & [-3.31,\ 0.774] \\
\bottomrule
\end{tabular}
\end{table}

Our final synthetic-data finding carries over to real-surface calibration: when the inverse network is trained primarily on GA-history data generated from the SPX target surface, it can produce parameter proposals whose implied prices track that same surface very closely. The figure below illustrates this effect for a representative strike slice: the co-evolutionary model-implied call prices align more tightly with market prices than the LHS-trained inverse model. However, given the earlier train--validation gap and the concentrated nature of optimizer-history sampling, this improvement is best interpreted as target-specific fitting rather than evidence of a uniformly better inverse mapping. In other words, GA-history training can yield superior in-sample replication of the target surface while sacrificing robustness away from that surface, whereas LHS provides broader coverage that typically improves out-of-sample stability.

\begin{figure}[H]
\centering
\includegraphics[width=\textwidth]{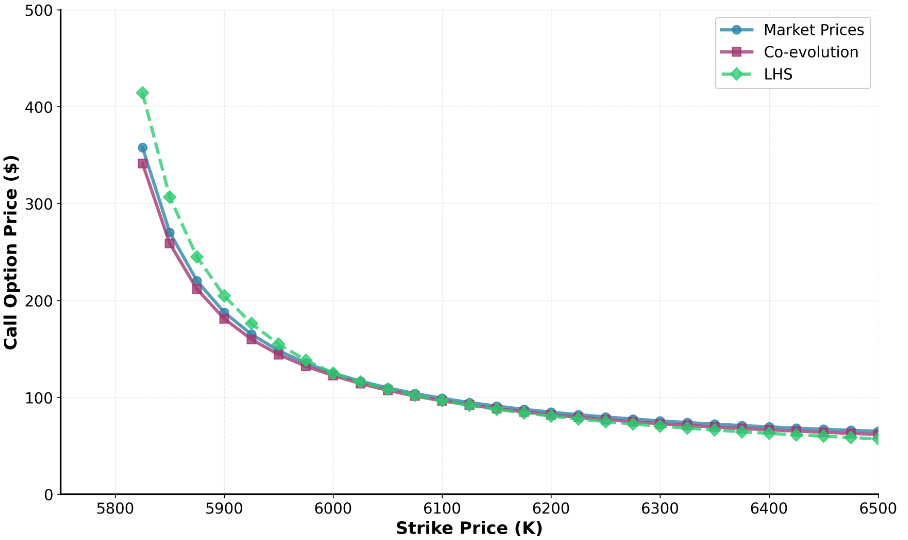}
\caption{Representative strike slice for SPX call prices comparing market quotes to model-implied prices produced by inverse networks trained with GA-history (co-evolutionary) data versus a space-filling LHS dataset.}
\label{fig:spx_slice_overfit}
\end{figure}

\section{Conclusion}
We proposed a dual-population co-evolutionary calibration framework for Heston in which a GA performs global parameter search while an evolving neural population learns an inverse mapping from option surfaces to parameters and provides warm-start seeds. Empirically, NN-driven injection accelerates GA convergence and sustains improvements beyond the point where a plain GA begins to plateau, and time-to-threshold experiments show that the evolutionary approach can reach LBFGS-level accuracy under the same pricing objective. At the same time, training inverse models exclusively on GA-history surfaces induces a concentrated data distribution and a widening train--validation gap, indicating overfitting to the target region. Comparisons against space-filling LHS datasets confirm that broader parameter-space coverage yields more stable and generalizable inverse mappings, even if it samples away from the target. In SPX calibration, GA-history-trained networks can reproduce the target surface more closely. However, due to the overfitting induced by optimizer-history sampling and the inherent ill-conditioning of Heston calibration, even small perturbations in the observed surface can map to parameter vectors far outside the training distribution, indicating that the apparent gains reflect target-specific fitting rather than a genuine improvement in learning a robust inverse mapping.

\bibliographystyle{plainnat}  
\bibliography{bib}            

@article{BlackScholes1973,
  author  = {Black, Fischer and Scholes, Myron},
  title   = {The Pricing of Options and Corporate Liabilities},
  journal = {Journal of Political Economy},
  year    = {1973},
  volume  = {81},
  number  = {3},
  pages   = {637--654}
}

@article{Heston1993,
  author  = {Heston, Steven L.},
  title   = {A Closed-Form Solution for Options with Stochastic Volatility with Applications to Bond and Currency Options},
  journal = {Review of Financial Studies},
  year    = {1993},
  volume  = {6},
  number  = {2},
  pages   = {327--343}
}

@article{CarrEtAl2002,
  author  = {Carr, Peter and Geman, Helyette and Madan, Dilip B. and Yor, Marc},
  title   = {The Fine Structure of Asset Returns: An Empirical Investigation},
  journal = {Journal of Business},
  year    = {2002},
  volume  = {75},
  number  = {2},
  pages   = {305--332}
}

@article{CoxIngersollRoss1985,
  author  = {Cox, John C. and Ingersoll, Jonathan E. and Ross, Stephen A.},
  title   = {A Theory of the Term Structure of Interest Rates},
  journal = {Econometrica},
  year    = {1985},
  volume  = {53},
  number  = {2},
  pages   = {385--407}
}

@article{Feller1951,
  author  = {Feller, William},
  title   = {Two Singular Diffusion Problems},
  journal = {Annals of Mathematics},
  year    = {1951},
  volume  = {54},
  number  = {1},
  pages   = {173--182}
}

@incollection{GilliSchumann2010,
  author    = {Gilli, Manfred and Schumann, Dietmar},
  title     = {Calibrating the Heston Model with Differential Evolution},
  booktitle = {Applications of Evolutionary Computation},
  editor    = {Di Chio, Cecilia and others},
  series    = {Lecture Notes in Computer Science},
  volume    = {6025},
  publisher = {Springer},
  year      = {2010},
  pages     = {242--250}
}

@incollection{GilliSchumann2011,
  author    = {Gilli, Manfred and Schumann, Dietmar},
  title     = {Calibrating Option Pricing Models with Heuristics},
  booktitle = {Natural Computing in Computational Finance},
  editor    = {Brabazon, Anthony and O'Neill, Michael},
  publisher = {Springer},
  year      = {2011},
  volume    = {4},
  pages     = {9--37}
}

@article{MugandaKasamani2023,
  author  = {Muganda, Ben W. and Kasamani, Betty S.},
  title   = {Parallelization and Acceleration of Dynamic Option Pricing Models on GPU--CPU Systems},
  journal = {Journal of Systems Science and Information},
  year    = {2023},
  volume  = {11},
  number  = {5},
  pages   = {622--635}
}

@article{OrtizEtAl2022,
  author  = {Ortiz, Alcides and others},
  title   = {Parameter Calibration of Heston's Model with Various Loss Functions and Methods},
  journal = {Contadur{\'\i}a y Administraci{\'o}n},
  year    = {2022},
  volume  = {67},
  number  = {1},
  pages   = {40--67}
}

@article{HaringHochreiter2015,
  author  = {Haring, Stefan and Hochreiter, Ronald},
  title   = {Efficient and Robust Calibration of the Heston Option Pricing Model for American Options Using an Improved Cuckoo Search Algorithm},
  journal = {Working Paper},
  year    = {2015}
}

@article{CuiEtAl2017,
  author  = {Cui, Yuhui and del Ba{\~n}o Rollin, Sergio and Germano, Guido},
  title   = {Full and Fast Calibration of the Heston Stochastic Volatility Model},
  journal = {European Journal of Operational Research},
  year    = {2017},
  volume  = {263},
  number  = {2},
  pages   = {625--638}
}

@article{LiuEtAl2019,
  author  = {Liu, Shuaiqi and Borovykh, Anastasia and Grzelak, Lech A. and Oosterlee, Cornelis W.},
  title   = {A Neural Network--Based Framework for Financial Model Calibration},
  journal = {Journal of Mathematics in Industry},
  year    = {2019},
  volume  = {9},
  number  = {9},
  pages   = {1--30}
}

@article{HorvathEtAl2021,
  author  = {Horvath, Blanka and Muguruza, Ander and Tomas, Mehdi},
  title   = {Deep Learning Volatility: A Deep Neural Network Perspective on Pricing and Calibration in (Rough) Volatility Models},
  journal = {Quantitative Finance},
  year    = {2021},
  volume  = {21},
  number  = {1},
  pages   = {11--27}
}

@article{BuchelEtAl2022,
  author  = {B{\"u}chel, Patrick and Kratochwil, Michael and Nagl, Maximilian and R{\"o}sch, Daniel},
  title   = {Deep Calibration of Financial Models: Turning Theory into Practice},
  journal = {Review of Derivatives Research},
  year    = {2022},
  volume  = {25},
  number  = {2},
  pages   = {109--136}
}

@article{ZhangEtAl2024,
  author  = {Zhang, Chen and Amici, Giovanni and Morandotti, Marco},
  title   = {Calibrating the Heston Model with Deep Differential Networks},
  journal = {arXiv preprint arXiv:2407.15536},
  year    = {2024}
}

\end{document}